\documentclass[11pt]{article}
\usepackage[a4paper]{geometry}
\usepackage{amsmath}
\usepackage{amsfonts}
\usepackage{amsbsy}
\usepackage{amsthm}
\usepackage{accents}
\usepackage{tensor}
\usepackage{authblk}
\usepackage{braket}
\usepackage{graphicx}
\usepackage{caption}
\usepackage{subcaption}
\usepackage{cite}
\newcommand{\bs}[1]{\mathbf{#1}}

\theoremstyle{definition}

\numberwithin{equation}{section}

\begin{document}
\title{\bf Note on rotational properties of position operators of massless particles}  
\author[1,2]{Michał~Dobrski\thanks{michal.dobrski@p.lodz.pl}}
\affil[1]{\small Institute of Physics, Lodz University of Technology,\linebreak Wólczańska 217/221, 93-005 Łódź, Poland}
\affil[2]{\small  Faculty of Physics and Applied Informatics, University of Lodz, \linebreak Pomorska 149/153, 90-236 Łódź, Poland} 
\date{}
\maketitle 
\begin{abstract}
Nonlinear action of the group of spatial rotations on commuting components of a~position operator of a massless particle of arbitrary helicity is studied. It is shown that linearization of this action necessarily leads to the Pryce operator with non-commuting components. The problem is also analyzed from a geometric perspective using Callan, Coleman, Wess and Zumino method.
%For a massless particle nonlinear action of the group of spatial rotations on a position operator with commuting components (Hawton operator) is studied. Using Callan, Coleman, Wess and Zumino method it is shown that coordinates yielding linearization of this action correspond to the Pryce operator with non-commuting components. 
\end{abstract}
\section{Introduction} 
%Contrary to the momentum and the angular momentum operators, a position operator is not defined by a representation of the Lie algebra of the spacetime symmetry group, but rather belongs to its universal enveloping algebra. 
The problem of definition of a position operator for relativistic particles has been studied since the works of Pryce \cite{pryce} and Newton and Wigner \cite{newton}. One of the important conclusions of these pioneering papers -- confirmed and extended over the years by further authors  \cite{wightman, skagerstam, kosinski} -- is that one cannot construct a position operator with all intuitively postulated properties for a typical massless particle e.g.\ for a photon. The precise version of this corollary was given by Skagerstam (who attributed it to Mukunda) in Ref.\ \cite{skagerstam}: for particles of helicity $|\lambda|>1/2$ there cannot exist a position operator which simultaneously has commuting components, and which transforms as a three-vector under spatial rotations.  
%
%One can argue that both these postulates are quite natural. The spacetime symmetry groups are defined by their action on the space and time coordinates. On the operational level the space coordinates can be, in principle, defined as positions of heavy particles forming the physical reference frame. Therefore, the transformation (and commutativity) properties of position operator under Euclidean group should coincide with that of Cartesian coordinates. For massless particles this appears to be impossible.

Thus, one must abandon either of the two conflicting assumptions. The historically first approach -- called the Pryce operator within this text -- favored vectorial transformation property over commutativity. On the other hand, in Ref. \cite{hawton} Hawton constructed a photon position operator whose components commute, but they do not transform as components of a three-vector. 
(It is well known that the question of the construction of the position operator for massless particles is very closely related to the issue of localizability. Although these matters are of key physical importance -- compare e.g.\ Ref. \cite{bialynicki-rev} -- we do not focus on them in the present work).

The Hawton operator was analyzed from various perspectives by several authors \cite{hawton2, hawtonbaylis, hawtonbaylis2, hawtondebierre, babaei, przanowski, przanowski2,jadczyk}. 
%However, the relation between Hawton and Pryce operators is usually described  by a simple statement that the first one involves a ``correction'' added to the latter. In this short paper we want to somewhat deepen this understanding 
In particular, in Ref. \cite{przanowski} a geometry corresponding to the relation between Hawton and Pryce operators was considered, and a ``correction'' that the first one adds to the latter was identified as a term which, at a cost of introducing singularities, flattens a connection corresponding to the Pryce operator. 

The aim of the present paper is to understand, for massless particles of arbitrary helicity, an ascpect of this relation associated with transformations under rotations.
We start (section 2) with an analysis of the action of the rotation group for the case of a position operator with commuting components. The corresponding nonlinearity (together with typical singularities) apear explicitly in the derived formulas for infinitesimal and finite rotations. The linearization problem is considered then -- we look for modifications of a position operator with commuting components which lead to the vectorial transformation law and conclude that the outcome, up to rather trivial corrections, must be the Pryce operator. The independence of this result from a particular choice of the angular momentum generators is discussed. Then (section 3) we look at the problem from a geometric point of view. The nonlinear transformation law is realized on a classical phase space and the group theoretic technique developed by Callan, Coleman, Wess and Zumino \cite{cwz, cwz2} is applied. In turn, the standard coordinates are constructed and the linearization result is rederived in the classical setting for one particular choice of the action of rotation generators. Finally (section 4), some concluding remarks are given.

Natural units, specifically $\hbar=1$, are used in this paper.

%\section{Standard coordinates for commuting $X_j$}
\section{Position operator for a massless particle and rotations}
Consider canonical commutation relations
\begin{equation}
\label{cano}
[X_k,p_l]=i   \delta_{kl}
\end{equation}
together with
\begin{equation}
\label{comm}
[X_k , X_l]=0
\end{equation}
and
\begin{equation}
[p_k,p_l]=0
\end{equation}
for $k,l = 1,2,3$. Thus, $X_k$ are commuting components of a canonical position operator. 
For helicity $|\lambda| >\frac{1}{2}$ these relations cannot be reconciled (see Ref.\ \cite{skagerstam, kosinski}) with three-vector transformation law 
\begin{equation}
\label{linact}
[J_k, X_l]=i \varepsilon_{klm} X_m.
\end{equation}

The following simple argument supporting this claim was given in Ref.\ \cite{skagerstam}. For the orbital angular momentum $L_k = \varepsilon_{klm} X_l p_m$ define an operator $S_k = J_k - L_k$. When the relations (\ref{cano}), (\ref{comm}) and (\ref{linact}) are altogether fulfilled, then $S_k$ has all the crucial properties of the spin. In turn a spectrum of the helicity operator must be given by all the numbers $|\lambda|, |\lambda| -1 , \dots , -|\lambda|+1, -|\lambda|$ and not only by $\lambda$ as one expects for a massless particle. There are therefore two possibilities. The first one is to discard the relation (\ref{comm}). Then one obtains the Pryce position operator whose components transform according to the three-vector rule (\ref{linact}), but which are non-commutative. The other is to abandon the rule (\ref{linact}), and a resulting position operator (which could be the Hawton operator in the case of photon) is then not a three-vector. As the first step, let us examine the transformation law corresponding to the action of the rotation group for this latter approach.

\subsection{Action of the rotation group}
Since the law being sought is dictated by the form of generators $J_k$, the corresponding massless representation of Poincar\'e algebra of helicity $\lambda$ should be introduced first. To this end we supplement $p_l$ with $H = |\vec{p}|$,
\begin{align}
J_1&= L_1 + \frac{  \lambda p_1}{|\vec{p}| +p_3}, &
J_2&= L_2 + \frac{  \lambda p_2}{|\vec{p}| +p_3}, &
J_3&= L_3 +   \lambda , 
\end{align} 
and
\begin{align}
\label{boosts}
N_1&= K_1 + \frac{  \lambda p_2}{|\vec{p}| +p_3}, &
N_2&= K_2 - \frac{  \lambda p_1}{|\vec{p}| +p_3}, &
N_3&= K_3 , 
\end{align} 
where $L_k = \varepsilon_{klm} X_l p_m$ and $K_l=|\vec{p}| X_l + \frac{i}{2} \frac{ p_l}{|\vec{p}|}$. Then, all the Poincar\'e algebra commutation relations 
\begin{align*}
[J_i,J_j]&=i\varepsilon_{ijk}J_k, &
[J_i,p_j]&=i\varepsilon_{ijk}p_k, &
[J_i,N_j]&=i\varepsilon_{ijk}N_k,\\
[N_i,N_j]&=-i\varepsilon_{ijk}J_k, &
[N_i,p_j]&=i\delta_{ij}H, &
[N_i,H]&=ip_i,\\
& & [J_i,H]=[p_i,H]&=[p_i,p_j]=0, & &
\end{align*}
can be easily verified. Nevertheless, it may look somewhat unusual to consider the representation -- which is an adapted version of that in Ref. \cite{novozhilov} (compare also Ref. \cite{lomont,frondsal}) -- written in terms of otherwise unspecified operators $X_k$. The advantage of this approach is that our discussion remains rooted in commutation properties and it is then applicable to various specific realizations of position operator with commuting components. For example $X_k$ may be an object acting on one-component wave functions as resulting from Wigner theory, or some Hawton operator defined for three-component wave functions. (However, in the latter case one needs to consider each helicity separately. This is possible as long as $X_k$ commutes with the helicity operator, which is a usual property of a Hawton operator, compare e.g.\ Ref.\ \cite{przanowski2}). Similarly, any particular definition of a scalar product remains irrelevant -- all the generators are Hermitian if the hermicity of $p_l$ and $X_k$ is guaranteed.

%%To obtain explicit formulas for $[J_k, X_l]$ one needs to settle $J_k$. 
%%The form of $J_k$ for massles particles results from the Wigner theory of representations of Poincar\'e group \cite{wigner}. 
%%(Although it is not unique due to ambiguity in the choice of a ``standard four-momentum'' $k$ and corresponding ``standard boost'' $L(p)$). 
%The particular realization of $J_k$ compatible with Wigner's theory of representations of Poincar\'e group \cite{wigner} can be chosen as 
%%(compare Ref.\ \cite{lomont, novozhilov,frondsal} and also Ref.\ \cite{hawtonbaylis})
%\begin{align}
%J_1&= L_1 + \frac{  \lambda p_1}{|\vec{p}| +p_3}, &
%J_2&= L_2 + \frac{  \lambda p_2}{|\vec{p}| +p_3}, &
%J_3&= L_3 +   \lambda , 
%\end{align} 
%where again $L_k = \varepsilon_{klm} X_l p_m$. (Compare Ref.\ \cite{lomont, novozhilov,frondsal} and also section II of Ref.\ \cite{hawtonbaylis}. Notice that all these references use operator  $i   \frac{\partial}{\partial p_k}$ in place of $X_k$. We can safely use the latter in the present context as it has the same commutation properties as $i   \frac{\partial}{\partial p_k}$). 
With the above choice of representation we are led to the following commutators
\begin{subequations}
\label{nonlinact}
\begin{align}
\nonumber
[J_1,X_1]&=- \frac{i \lambda  }{|\vec{p}| (1+\kappa_3)^2} (1-\kappa_1^2 +\kappa_3), &
[J_2,X_1]&=-i   X_3 + \frac{i \lambda   \kappa_1 \kappa_2}{|\vec{p}| (1+\kappa_3)^2} , \\
\nonumber
[J_1,X_2]&=i   X_3 + \frac{i \lambda   \kappa_1 \kappa_2}{|\vec{p}| (1+\kappa_3)^2}, &
[J_2,X_2]&=- \frac{i \lambda  }{|\vec{p}| (1+\kappa_3)^2} (1-\kappa_2^2 +\kappa_3),\\
\label{nonv}
[J_1,X_3]&= - i   X_2 + \frac{i \lambda   \kappa_1 }{|\vec{p}| (1+\kappa_3)}, &
[J_2,X_3]&=i   X_1 + \frac{i \lambda   \kappa_2 }{|\vec{p}| (1+\kappa_3)},
\end{align}
and
\begin{equation}
[J_3,X_m]=i   \varepsilon_{3mk} X_k,
\end{equation}
\end{subequations}
for $\kappa_i=\frac{p_i}{|\vec{p}|}$, which can be summarized in a formula
\begin{equation}
[J_k,X_l]=i\varepsilon_{klm}X_m +R_{kl}
\end{equation}
with $R_{kl}$ defined by right hand sides of relations (\ref{nonlinact}).

As an example of a finite transformation corresponding to commutators (\ref{nonlinact}) let us consider a rotation about the first axis by an angle $\theta$. Integrating infinitesimal transformations given by (\ref{nonv}) one obtains
\begin{align}
\nonumber
X_1 \to X_1 +\frac{  \lambda}{|\vec{p}|} \frac{\left(\kappa_2 - \kappa_2' +(1-\kappa_1^2)\sin\theta\right)}{(1+\kappa_3)(1+\kappa_3')}\\
\nonumber
X_2 \to X_2 \cos\theta - X_3 \sin\theta + \frac{  \lambda \kappa_1}{|\vec{p}|}  \left(\frac{1}{1+\kappa_3'} - \frac{\cos \theta}{1+\kappa_3} \right)\\
\label{nonlinfinit}
X_3 \to X_2 \sin\theta + X_3 \cos\theta - \frac{  \lambda}{|\vec{p}|} \frac{\kappa_1 \sin \theta}{1+ \kappa_3}
\end{align}
where $\kappa'_j$ denotes components of $\vec{\kappa}$ rotated about first axis by an angle $\theta$ in the usual (vectorial) way. Notice that the above rule is not well defined if either $\vec{\kappa}$ or $\vec{\kappa}'$ makes relations (\ref{nonv}) singular.

\subsection{Linearization problem}
\label{quantlinsect}
The problem we are going to address now is how to construct  -- starting from the operator $X_k$ -- a new position operator $Y_k$, which would transform in a vectorial way under rotations.
Since operator $Y_k$ should satisfy the canonical relations (\ref{cano}) it must be of the form 
\begin{equation}
Y_k = X_k + f_k(\vec{p}).
\end{equation}
Then, from
\begin{equation}
\label{simplelincond}
[J_k,Y_l] = i \varepsilon_{klm}Y_m
\end{equation}
one obtains the system of linear differential equations for $f_k$
\begin{equation}
\label{nhmg}
 \frac{\partial f_l}{\partial p_m}\varepsilon_{kmn}  p_n - \varepsilon_{klm} f_m = i R_{kl}.
\end{equation}
The general solution of its homogeneous part
\begin{equation}
\label{fhmg}
 \frac{\partial f^{(0)}_l}{\partial p_m}\varepsilon_{kmn}  p_n - \varepsilon_{klm} f^{(0)}_m = 0
\end{equation}
can be easily obtained\footnote{Taking $k=l$ in equations (\ref{fhmg}) one derives $f_l=f_l(|\vec{p}|^2 - p_l^2,p_l)$. This produces relation $\varepsilon_{jmn}f_m p_n=0$ from the remaining equations. In turn, the system (\ref{fhmg}) can be decoupled and the final result can be obtained.}, yielding the expected result $f^{(0)}_k = \varphi(|\vec{p}|) p_k$. Hence, it remains to point out any particular solution of the system (\ref{nhmg}). One way of doing this is by guessing it from relations (\ref{nonlinact}). Indeed, if there exists a solution such that $Y_3=X_3$  then $Y_1$ and $Y_2$ can be extracted from equations (\ref{nonv}) that involve $[J_2,X_3]$ and $[J_1,X_3]$ respectively. This yields
\begin{align}
\label{specialsol}
f_1&= \frac{ \lambda   \kappa_2 }{|\vec{p}| (1+\kappa_3)}, &
f_2&= - \frac{ \lambda   \kappa_1 }{|\vec{p}| (1+\kappa_3)}, &
f_3&= 0,
\end{align}
which can be verified to be a valid solution of the system (\ref{nhmg}). Thus the general solution of the linearization problem reads
\begin{align}
\label{solquant}
Y_1  &=  X_1 +\frac{\lambda \kappa_2}{|\vec{p}|(1+\kappa_3)} + \varphi p_1, &
Y_2  &=  X_2 -\ \frac{\lambda \kappa_1}{|\vec{p}|(1+\kappa_3)} +\varphi p_2, &
Y_3  &= X_3 + \varphi p_3.
\end{align}
where $\varphi=\varphi(|\vec{p}|)$ is an arbitrary function of $|\vec{p}|$. The straightforward calculation shows that components of position operator $Y_k$ do not commute
\begin{equation}
\label{simplecomm}
[Y_i,Y_j]=-\frac{i \lambda}{|\vec{p}|^3} \varepsilon_{ijk} p_k.
\end{equation} 
Hence, we have obtained a standard commutation relations of a vectorial, non-commuting position operator (compare e.g.~Ref.\ \cite{skagerstam,kosinski}) of a massless particle with helicity $\lambda$. Indeed, using  boosts $N_k$ defined by formula (\ref{boosts}) one can rewrite the solution (\ref{solquant}) explicitly as a Pryce operator 
\begin{equation}
\label{yintermsofn}
Y_k =\frac{1}{2} \left( \frac{1}{|\vec{p}|} N_k + N_k \frac{1}{|\vec{p}|} \right) + \varphi p_k.
\end{equation}
Moreover, it can be observed that for $\lambda=\pm 1$ and $\varphi \equiv 0$ the expressions (\ref{solquant}) reproduce the relation between Pryce operator and a variant of Hawton operator (see Ref.\ \cite{hawtonbaylis}, and also the formula (2.67) in Ref.\ \cite{przanowski}).

We should comment on the dependence of the presented construction on a choice of a massless representation of Poincar\'e algebra. Consider an unitary operator $U(\vec{p})$ depending solely on $\vec{p}$. Then, generators $p_k$ and $H$  remain unchanged under an automorphism corresponding to $U(\vec{p})$, while $J_k$ and $N_l$ transform according to the rule
\begin{align}
\tilde{J}_k &= U(\vec{p}) J_k U^{-1}(\vec{p}) , &  \tilde{N}_l &= U(\vec{p}) N_l U^{-1}(\vec{p}).
\end{align}
One can use  $U(\vec{p})$ to bring angular momentum generators to any desired form allowed for a massless particle of helicity $\lambda$. For example, taking
%\begin{equation}
%U(\vec{p}) = \Bigg(\frac{p_2^2
%   (p_1+p_3)}{\left(|\vec{p}|^2 - p_1^2\right)
%   \left(|\vec{p}|+p_3\right)}+\frac{p_3^2}{|\vec{p}|^2 - p_1^2}+ i \frac{ p_2
%   \left(|\vec{p}|
%   \left(|\vec{p}|+p_3\right)-p_1
%   (p_1+p_3)\right)}{\left(|\vec{p}|^2 - p_1^2\right)
%  \left(|\vec{p}|+p_3\right)}\Bigg)^\lambda ,
%\end{equation}
\begin{equation}
U(\vec{p}) = \Bigg(\frac{\kappa_2^2
   (\kappa_1+\kappa_3)}{\left(1 - \kappa_1^2\right)
   \left(1+\kappa_3\right)}+\frac{\kappa_3^2}{1 - \kappa_1^2}+ i \frac{ \kappa_2
   \left(
   1+\kappa_3-\kappa_1
   (\kappa_1+\kappa_3)\right)}{\left(1 - \kappa_1^2\right)
   \left(1+\kappa_3\right)}\Bigg)^\lambda ,
\end{equation}
with $\lambda \in \mathbb{Z}$,  one obtains  Lomont-Moses \cite{lomont} form of $\tilde{J}_k$
\begin{align}
\tilde{J}_1&= L_1 + \lambda , &
\tilde{J}_2&= L_2 + \frac{  \lambda p_2}{|\vec{p}| +p_1}, &
\tilde{J}_3&= L_3 +   \frac{  \lambda p_3}{|\vec{p}| +p_1}  . 
\end{align} 
But since operators $Y_k$ satisfy condition (\ref{simplelincond}) then 
\begin{equation}
\tilde{Y}_k =U(\vec{p}) Y_k U^{-1}(\vec{p}) =  Y_k + i U(\vec{p}) \frac{\partial U ^{-1}(\vec{p})}{\partial p_k}
\end{equation}
must fulfill relation
\begin{equation}
[\tilde{J}_k,\tilde{Y}_l] = i \varepsilon_{klm} \tilde{Y}_m.
\end{equation}
Thus $\tilde{Y}_k $ solve the linearization problem formulated in terms of $\tilde{J}_k$. It follows immediately, that the commutator $[\tilde{Y}_i,\tilde{Y}_j]$ is again given by r.h.s.\ of the formula  (\ref{simplecomm}), and that $\tilde{Y}_k$ can be expressed in terms of $\tilde{N}_k$ as in the relation (\ref{yintermsofn}).

%\section{Actions of the rotation group on a classical phase space}
\section{Geometric approach}
We are going to look at the problem of linearization from a geometric perspective by reformulating it on a classical phase space. Eventually, a counterpart of the special solution (\ref{specialsol}) guessed in the previous section will be constructed by a systematic procedure.  
Nonlinear classical realization of the rotation group obtained from the relations (\ref{nonlinact}) will be used for this purpose. The group, $SU(2)$, is compact and simple, so all its realizations can be investigated using the method developed by Callan, Coleman, Wess and Zumino, which will be briefly recalled now.

\subsection{Method of nonlinear realizations}
The linearity of quantum mechanics on the level of the space of states does not rule out the nonlinear character of the transformation properties of certain quantities (in our case -- particular elements of algebra of observables) under the action of a given symmetry group. Originally this idea appeared to be very fruitful in low energy meson physics and has been put into an elegant general framework by Callan, Coleman, Wess and Zumino in Ref.\ \cite{cwz,cwz2}. (See e.g. Ref. \cite{hammer, brivio} for some further examples of applications in effective field theories). 

Consider compact connected semi-simple Lie group $G$ acting on a given manifold $M$, and denote this action by $g\circ m$ for $g\in G$ and $m \in M$. Let $V_i$, $A_l$ form a complete orthonormal set of generators of $G$, with $V_i$ being generators of a subgroup $H$ of the group $G$. An arbitrary element $g \in G$ can be uniquely decomposed as
\begin{equation}
g=e^{\xi_l A_l} e^{u_i V_i}.
\end{equation}
This also applies to $g_0 e^{\xi_l A_l}$ for any $g_0 \in G$, i.e.
\begin{equation}
\label{xiprimedef}
g_0 e^{\xi_l A_l} = e^{\xi'_l(\xi,g_0) A_l} e^{u'_i(\xi,g_0) V_i},
\end{equation}
where the functions $\xi'_l(\xi,g_0)$ and $u'_i(\xi,g_0)$ result solely from the structure of the group (and the choice of generators $V_i$, $A_l$). Suppose that $H$ is a \emph{stability subgroup} of some point $m_0 \in M$, i.e.\ $h\circ m_0=m_0$ for all $h\in H$. The algebraic part of the core result of Ref.\ \cite{cwz} can be summarized as follows. It is always possible to introduce on $M$ coordinates centered at $m_0$ and denoted by $(\xi,\psi)$ (the \emph{standard coordinates}) in such a way, that the action of $G$ on $M$ takes form
\begin{equation}
\label{transrule}
g \circ (\xi,\psi)=(\xi',\psi'),
\end{equation}
with $\xi'$ given precisely by the group transformation rule (\ref{xiprimedef}), and
\begin{equation}
\label{psirule}
\psi' = D(e^{u'_i V_i}) \psi,
\end{equation}
where $D$ is a linear representation of $H$, while $u'$ is also defined by formula (\ref{xiprimedef}). By the very construction $\xi$ coordinates can be obtained as quantities parametrizing orbit through $m_0$ (or corresponding to the ``transitive part''  of the action of $G$). On the other hand if manifold $M$ is (locally) parametrized by $(\xi,y)$ a transition to a standard coordinates $(\xi,\psi)$ can be achieved (compare Ref.\ \cite{cwz2}) by computing $\psi$ defined by the relation
\begin{equation}
\label{ytopsirule}
e^{-\xi_l A_l} \circ (\xi,y) = (0, \psi).
\end{equation}

It can be observed that for an element $h \in H$ the relation (\ref{transrule}) becomes linear
\begin{equation}
\label{hsublincase}
h \circ (\xi,\psi)=(D^{(b)}(h) \xi', D(h) \psi'),
\end{equation} 
where $D^{(b)}$ is some linear representation of $H$. However, for a general element $g \in G$, even the rule (\ref{psirule}) is nonlinear due to the dependence of $u'$ on $\xi$. One can bring this nonlinear relation to a completely linear one (with respect to the whole group $G$) by an appropriate redefinition of variables. It is enough to consider representation $D$ of $G$ such that when restricted to subgroup $H$ it coincides with representation appearing in (\ref{psirule}). Then the quantities introduced as
\begin{equation}
\label{psitofrule}
f(\xi,\psi) = D(e^{\xi_l A_l}) \psi
\end{equation}
transform in a fully linear way (compare Ref.\ \cite{cwz} for details and a more general approach).

\subsection{Standard coordinates for a classical phase space}
Let us apply the above scheme to study a classical variant of the action of the rotation group given by relations (\ref{nonlinact}). Therefore we consider generators of $SU(2)$ written in terms of Pauli matrices
\begin{align}
A_1 &=  -\frac{i}{2} \sigma_{1}, & A_2 &= -\frac{i}{2} \sigma_{2}, & V &= -\frac{i}{2} \sigma_{3},     
\end{align}
whose infinitesimal action on a phase space parametrized by classical quantities $(p_j,X_k)$ is described by the vector fields
\begin{align}
\nonumber
\mathcal{A}_1  = &  -p_3 \frac{\partial}{\partial p_2} + p_2 \frac{\partial}{\partial p_3} + \frac{\lambda(1 - \kappa_1^2 + \kappa_3)}{|\vec{p}| (1 +\kappa_3)^2} \frac{\partial}{\partial X_1}\\
\nonumber
&-
 \left(X_3  +  \frac{\lambda \kappa_1 \kappa_2}{|\vec{p}| (1 +\kappa_3)^2} \right)\frac{\partial}{\partial X_2}+
\left( X_2 - \frac{\lambda \kappa_1}{|\vec{p}| (1 + \kappa_3)} \right) \frac{\partial}{\partial X_3}
 ,
\\
\nonumber
\mathcal{A}_2  = &  + p_3\frac{\partial}{\partial p_1}
-p_1\frac{\partial}{\partial p_3}+
\left(X_3 - \frac{\lambda \kappa_1 \kappa_2}{|\vec{p}| (1 +\kappa_3)^2} \right) \frac{\partial}{\partial X_1}\\
\nonumber
 & +
\frac{ \lambda (1 -\kappa_2^2 + \kappa_3)}{|\vec{p}| (1 +\kappa_3)^2} \frac{\partial}{\partial X_2}
-\left(X_1 + \frac{\lambda \kappa_2}{|\vec{p}| (1 + \kappa_3)} \right) \frac{\partial}{\partial X_3},
\\
\label{vecfields}
\mathcal{V}  = &   -p_2\frac{\partial}{\partial p_1} +p_1\frac{\partial}{\partial p_2} -X_2 \frac{\partial}{\partial X_1} + X_1 \frac{\partial}{\partial X_2}
\end{align} 
respectively, obtained from relations (\ref{nonlinact}) and from $[J_i,p_j]=i\varepsilon_{ijk}p_k$.

In the first step we look for variables $\xi_\alpha$ expressed in terms of momentum variables $p_k$. Clearly, an arbitrary $g \in SU(2)$ acts on coordinates $p_j$ by the usual three-vector transformation law, which can be described by the representation
\begin{equation}
\label{stdrep}
\vec{p} \cdot \vec{\sigma} \to g (\vec{p} \cdot \vec{\sigma}) g^\dagger = \vec{p}^{\; \prime} \cdot \vec{\sigma}.
\end{equation}
The stability subgroup of any nonzero momentum vector, which we choose to be $\vec{p}_0=(0,0,|\vec{p}|)$, is $U(1)$. Setting $m_0=(\vec{p}_0,0)$ we obtain that the corresponding orbit $S^2$ is parametrized by $\xi_\alpha$, $\alpha=1,2$. (This parametrization holds to be smooth in a~neigbourhood of  $\vec{p}_0$ only). Explicitly, we put
\begin{equation}
\label{eqpk}
\vec{p} \cdot \vec{\sigma} =e^{\xi_\beta A_\beta} |\vec{p}| \sigma_3 (e^{\xi_\beta A_\beta} )^\dagger
\end{equation} 
with $A_\beta=-\frac{i}{2} \sigma_\beta$, $\beta=1,2$. From (\ref{eqpk}) one easily infers
%one easily infers
\begin{align}
\label{xitop}
\frac{p_\alpha}{|\vec{p}|} &=  \frac {\sin |\xi| }{|\xi|} \varepsilon_{\alpha \beta} \xi_\beta  & \frac{p_3}{|\vec{p}|} &= \cos|\xi|
\end{align}
where $|\xi|=\sqrt{\xi_1^2+ \xi_2^2 }$, while $\varepsilon_{\alpha \beta}$ is the two-dimensional Levi-Civita symbol and $\alpha, \beta=1,2$ with summation over repeated index applied. Notice that the remaining generator $V = -\frac{i}{2} \sigma_3$ produces subgroup $U(1)$ which leaves $\xi_\alpha = 0$ invariant and also  yields a linear rotation of $X_k$ about third axis.
 
The transformation rule for $\xi_\alpha$ can be calculated from the following version of the formula (\ref{xiprimedef}) 
%In the similar way one can obtain $\xi_\alpha'(\xi,g_0)$ and $u'(\xi,g_0)$ defined by
\begin{equation}
\label{updef}
g_0 e^{\xi_\beta A_\beta} = e^{\xi_\beta' (\xi,g_0) A_\beta} e^{u'(\xi,g_0) V}
\end{equation}
where $g_0 \in SU(2)$. The relations obtained for infinitesimal $g_0=\bs{1}+ \tau A_1$ read
\begin{subequations}
\label{xiutransf}
\begin{align}
\nonumber
\xi_1'(\xi,\bs{1}+ \tau A_1) &= \xi_1 + \tau \frac{\xi_1^2 + \xi_2^2 |\xi| \cot |\xi|}{|\xi|^2}, \\
\nonumber
\xi_2'(\xi,\bs{1}+ \tau A_1) &= \xi_2 + \tau \frac{\xi_1 \xi_2 (1- |\xi| \cot |\xi|)}{|\xi|^2},\\
u'(\xi,\bs{1}+ \tau A_1) &= \tau \frac{\xi_2}{|\xi|} \tan \left(\frac{|\xi|}{2}\right),
\end{align}
and
\begin{align}
\nonumber
\xi_1'(\xi,\bs{1}+ \tau A_2) &= \xi_1 + \tau \frac{\xi_1 \xi_2 (1- |\xi| \cot |\xi|)}{|\xi|^2},\\
\nonumber
\xi_2'(\xi,\bs{1}+ \tau A_2) &= \xi_2 + \tau \frac{\xi_2^2 + \xi_1^2 |\xi| \cot |\xi|}{|\xi|^2},\\
u'(\xi,\bs{1}+ \tau A_2) &= -\tau \frac{\xi_1}{|\xi|} \tan \left(\frac{|\xi|}{2}\right),
\end{align}
for $g_0=\bs{1}+ \tau A_2$. Consistently with formula (\ref{hsublincase}), the rotation about third axis results in
\begin{align}
\nonumber
\xi_1'(\xi,e^{\tau V}) &= \cos (\tau) \xi_1 -\sin(\tau) \xi_2,\\
\nonumber
\xi_2'(\xi,e^{\tau V}) &= \sin(\tau) \xi_1 + \cos(\tau) \xi_2,\\
u'(\xi,e^{\tau V}) &= \tau.
\end{align}
\end{subequations}

Quantities $(\xi_1,\xi_2)$ form the nonlinear realization of $SU(2)$ group. 
%In the terminology of Ref.\ \cite{cwz} they are the preferred (Goldstone) coordinates. 
In our case they correspond to momentum variables (cf. equation (\ref{xitop})). Now we need to complete $(\xi_1,\xi_2)$ by the additional variables $\psi$, the adjoint variables, which transform according to the rule (\ref{psirule}), i.e.
%The remaininig standard coordinates $\psi_j$ correspond to $X_j$ and should transform under $g_0 \in SU(2)$ according to
\begin{equation}
\label{psitransform}
\psi \to D(e^{u'(\xi,g_0) V}) \psi,
\end{equation}   
where $u'(\xi,g_0)$ is calculated from eq.\ (\ref{updef}) and $D$ is a linear representation of $U(1)$. Our manifold is parametrized by $(\xi_\alpha,X_i)$ so far. (For the sake of completeness one should also introduce a variable related to $|\vec{p}|$; however $|\vec{p}|$ transforms trivially and in what follows we keep $|\vec{p}|$ as a parameter). In order to put the parametrization in the standard form $(\xi_\alpha,\psi_i)$ one has to compute $\psi$ from the formula (\ref{ytopsirule})
%where $D$ is a linear representation of $U(1)$. (For the sake of completeness one could also introduce $\psi_0$ related to $|\vec{p}|$. This variable transforms trivially and in what follows we keep $|\vec{p}|$ as a parameter). One can find $\psi_j$ using the formula given in \cite{cwz2}: the point with coordinates $(\xi, X)$ is described by new coordinates $(\xi, \psi)$, where $\psi_j$ is calculated from the relation
\begin{equation}
e^{-\xi_\beta A_\beta} \circ (\xi,X) = (0, \psi),
\end{equation}
for $\circ$ denoting action of $SU(2)$. Since vector fields (\ref{vecfields}) describe this action infinitesimally it is convenient to introduce $\psi_{(t)}$ by
\begin{equation}
e^{-t \xi_\beta A_\beta} \circ (\xi,X) = ((1-t) \xi, \psi_{(t)}),
\end{equation}
and to observe that
\begin{equation}
((1-(t+\tau)) \xi, \psi_{(t+\tau)})=
%e^{-(t+\tau) \xi_\beta A_\beta} \circ (\xi,X)= 
e^{-\tau \xi_\beta A_\beta} \circ ((1-t) \xi, \psi_{(t)}).
\end{equation}
In turn 
\begin{equation}
\label{psieq}
%\dot{\psi}_{(t)} = \left. \delta_{-\xi_\beta A_\beta}  \psi_{(t)} \right|_{(1-t)\xi}
\dot{\psi}_{(t)} = \delta \psi_{(t)}, 
\end{equation}
where $\delta \psi_{(t)}$ denotes a transformation of $\psi_{(t)}$ under an action of the generator $-\xi_\beta A_\beta$, which should be calculated from formulas (\ref{vecfields}), with $\kappa_i$ determined by $(1-t) \xi_\alpha$. The initial condition for equation (\ref{psieq}) is $\psi_{j(0)}=X_j$ and the standard coordinates are $\psi_j =\psi_{j(1)}$. The explicit version of relation (\ref{psieq}) reads
\begin{align}
\nonumber
\dot{\psi}_{1 (t)} &= -\xi_2 \psi_{3 (t)} -  \frac{   \lambda\xi_1}{|\vec{p}| \left(1+ \cos\left((1-t) |\xi| \right)\right)}, \\
\nonumber
\dot{\psi}_{2 (t)} &= \xi_1 \psi_{3 (t)} -  \frac{   \lambda\xi_2}{|\vec{p}| \left(1+ \cos\left((1-t) |\xi| \right)\right)},\\
\label{psieq2}
\dot{\psi}_{3 (t)} &= \xi_2 \psi_{1 (t)} -\xi_1 \psi_{2 (t)}.
\end{align}
The solution at $t=1$ of this system of linear inhomogeneous equations is given by
\begin{align}
\nonumber
\psi_1  &= \tilde{\psi}_1  -\frac{  \lambda}{|\vec{p}|} \frac{\xi_1 \tan \frac{|\xi|}{2}}{ |\xi|} = \tilde{\psi}_1  +\frac{  \lambda}{|\vec{p}|} \frac{\kappa_2}{1+\kappa_3},\\
\nonumber
\psi_2  &= \tilde{\psi}_2  -\frac{  \lambda}{|\vec{p}|} \frac{\xi_2 \tan \frac{|\xi|}{2}}{ |\xi|} = \tilde{\psi}_2  -\frac{  \lambda}{|\vec{p}|} \frac{\kappa_1}{1+\kappa_3},\\
\psi_3  &= \tilde{\psi}_3, 
\end{align} 
where the functions
\begin{align}
\nonumber
\tilde{\psi}_1 &=\frac{1}{|\xi|^2}\Big( X_1 (\xi_1^2+ \xi_2^2 \cos|\xi|) + X_2 \xi_1 \xi_2 (1-\cos|\xi|) - X_3 \xi_2 |\xi| \sin |\xi|\Big)\\
&= X_1 \frac{\kappa_2^2 + \kappa_1^2 \kappa_3}{1-\kappa_3^2} - X_2 \frac{\kappa_1 \kappa_2 (1-\kappa_3)}{1-\kappa_3^2} - X_3 \kappa_1,\\
\nonumber
\tilde{\psi}_2  &=\frac{1}{|\xi|^2}\Big( X_1 \xi_1 \xi_2 (1-\cos|\xi|)  + X_2 (\xi_2^2+ \xi_1^2 \cos|\xi|) + X_3 \xi_1 |\xi| \sin |\xi|\Big)\\
&=-X_1 \frac{\kappa_1 \kappa_2 (1-\kappa_3)}{1-\kappa_3^2} +X_2  \frac{\kappa_1^2 + \kappa_2^2 \kappa_3}{1-\kappa_3^2} -X_3 \kappa_2,\\
\nonumber
\tilde{\psi}_3  &=\frac{\sin |\xi|}{|\xi|} \big( X_1 \xi_2 - X_2 \xi_1 \big)+ X_3 \cos|\xi|\\
&=X_1 \kappa_1 + X_3 \kappa_3 + X_3 \kappa_3
\end{align}
are the solution of a homogeneous variant of (\ref{psieq2}) taken at $t=1$. 
(It is probably worth to remark that homogeneous part of equations (\ref{psieq2}) corresponds to the transformation rule given by relation (\ref{linact}). Thus, $\tilde{\psi}_j$ would play the role of adjoint variables, if $X_j$ transformed as a standard three-vector).
 
\subsection{Linearization of group action}
Now we want to construct from $(\xi,\psi)$ new coordinates $Y_j$, for which the group action of $SU(2)$ becomes linear. To this end (cf.\ formula (\ref{psitofrule})) it is enough to take a three-dimensional representation $D$ of $SU(2)$ such that its restriction to $U(1)$ generated by $V$ is the representation appearing in (\ref{psitransform}), and to introduce
\begin{equation}
Y_j = D_{jk}(e^{\xi_\beta A_\beta})\psi_k.
\end{equation}
Using formulas (\ref{nonlinact}) and (\ref{xiutransf}) one can verify that the transformation (\ref{psitransform}) involves rotation about third axis and consequently the representation defined by relation (\ref{stdrep}) can be taken as $D$. In turn $Y_j$ calculated from
\begin{equation}
\vec{Y} \cdot \vec{\sigma} =e^{\xi_\beta A_\beta} (\vec{\psi} \cdot \vec{\sigma}) (e^{\xi_\beta A_\beta})^\dagger
\end{equation}
read
\begin{align}
\nonumber
Y_1  &= X_1 -\frac{ \lambda}{|\vec{p}|} \frac{\xi_1 \tan \frac{|\xi|}{2}}{ |\xi|} = X_1 +\frac{ \lambda}{|\vec{p}|} \frac{\kappa_2}{1+\kappa_3},\\
\nonumber
Y_2  &= X_2 -\frac{ \lambda}{|\vec{p}|} \frac{\xi_2 \tan \frac{|\xi|}{2}}{ |\xi|} = X_2 -\frac{ \lambda}{|\vec{p}|} \frac{\kappa_1}{1+\kappa_3},\\
\label{comnoncom}
Y_3  &= X_3.
\end{align}
%At the quantum level the commutator for these new variables is given by
%\begin{equation}
%[Y_i,Y_j]=-\frac{i \lambda}{|\vec{p}|^3} \varepsilon_{ijk} p_k.
%\end{equation}
Thus, in the present setting we obtain a classical variant of the special solution (\ref{specialsol}) and Pryce variables (\ref{solquant}) as a result of linearization.

\section{Final comments}
The general conclusion of the present paper is that, for a massless particle of nonzero helicity, the linearization of the corresponding action of the rotation group on commuting components of position operator necessarily leads to the Pryce operator. This result was shown to be independent from particular choice of a representation of Poincar\'e algebra.  The problem was also considered in a classical framework using Callan, Coleman, Wess and Zumino method and for a specific choice of the action of rotation group generators. The classical counterpart of the Pryce variables re-emerged in this differential geometric approach. 

\section*{Acknowledgments}
I would like to express my deep gratitude to P. Kosiński for proposing the research topic, explaining details of the method of nonlinear realizations and for all the kind help he has given me during and after the research internship I~had at the University of Lodz. I~am very grateful to A. Jadczyk for an enlightening discussion, especially for fixing the logic of section \ref{quantlinsect}.  The instructive comments of the reviewers helped to shape the final form of this work.


\begin{thebibliography}{999}
\hyphenation{Deut-scher}
\bibitem{pryce} M.~H.~L.~Pryce, \emph{Proc. R. Soc. London, Ser. A} \textbf{195}, 62 (1948)
\bibitem{newton} T.~D.~Newton and E.~P.~Wigner, \emph{Rev. Mod. Phys.} \textbf{21}, 400 (1949)
\bibitem{wightman} A.~S.~Wightman, \emph{Rev. Mod. Phys.} \textbf{34}, 845 (1962)
%\bibitem{jauch} J.~M~Jauch and C.~Piron, \emph{Helv. Phys. Acta} \textbf{40}, 559 (1967)
%\bibitem{amrein} W.~O.~Amrein, \emph{Helv. Phys. Acta} \textbf{42}, 149 (1969)
\bibitem{skagerstam} B.-S.~K. Skagerstam, \emph{Localization of massless spinning particles and the Berry phase} in ``On Klauder’s Path: A Field
 Trip'', pp. 209-222, Eds. G.~G. Emch and G.~C. Hegerfeldt, World Scientific, Singapore (1994).
\bibitem{kosinski} P.~Kosiński and P.~Maślanka, \emph{Ann. Phys.} \textbf{398}, 203 (2018)
\bibitem{hawton} M.~Hawton, \emph{Phys. Rev. A} \textbf{59}, 954 (1999)
\bibitem{bialynicki-rev} I.~Białynicki-Birula and Z.~Białynicka-Birula, \emph{Phys. Rev. A} \textbf{86}, 022118 (2012)
\bibitem{hawton2} M.~Hawton, \emph{Phys. Rev. A} \textbf{59}, 3223 (1999)
\bibitem{hawtonbaylis} M.~Hawton and W.~E. Baylis, \emph{Phys. Rev. A} \textbf{64}, 012101 (2001)
\bibitem{hawtonbaylis2} M.~Hawton and W.~E. Baylis, \emph{Phys. Rev. A} \textbf{71}, 033816 (2005)
\bibitem{hawtondebierre} M.~Hawton and V.~Debierre, \emph{J. Math. Phys.} \textbf{60}, 052104 (2019)
\bibitem{babaei} H.~Babaei and A.~Mostafazadeh, \emph{J. Math. Phys.} \textbf{58}, 082302 (2017)
\bibitem{przanowski} M.~Dobrski, M.~Przanowski, J.~Tosiek and F.~J.~Turrubiates, \emph{Phys. Rev. A} \textbf{104}, 042206 (2021)
\bibitem{przanowski2} M.~Dobrski, M.~Przanowski, J.~Tosiek and F.~J.~Turrubiates, \emph{Phys. Rev. A} \textbf{107}, 042208 (2023)
\bibitem{jadczyk} A. Jadczyk, \emph{Mathematics} \textbf{12}, 1140, (2024)
\bibitem{cwz} S.~Coleman, J.~Wess and B.~Zumino, \emph{Phys. Rev.} \textbf{177}, 2239 (1969)
\bibitem{cwz2} C.~G.~Callan, S.~Coleman, J.~Wess and B.~Zumino, \emph{Phys. Rev.} \textbf{177}, 2247 (1969)
\bibitem{wigner} E.~P.~Wigner, \emph{Ann. Math.} \textbf{40}, 149 (1939)
\bibitem{novozhilov} Y.~V.~Novozhilov \emph{Introduction to Elementary Particle Theory} (Pergamon Press, Oxford 1975)
\bibitem{lomont} J.~S.~Lomont and H.~E.~Moses, \emph{J. Math. Phys.} \textbf{3}, 405 (1962)
\bibitem{frondsal} C.~Fronsdal, \emph{Phys. Rev.} \textbf{113}, 1367 (1959)
\bibitem{hammer} H.-W.~Hammer, S.~König, and U.~van~Kolck, \emph{Rev. Mod. Phys.} \textbf{92}, 025004 (2020)
\bibitem{brivio} I.~Brivio and M.~Trott, \emph{Phys. Rep.} \textbf{793}, 1 (2019) 

\end{thebibliography}
\end{document}